\documentclass[aip,apl,reprint]{revtex4-1}
\usepackage{graphicx}

\newcommand{\nvsp}{NV$^-$ }

\newcommand{\nvsps}{NV$^-$s }
\draft % marks overfull lines with a black rule on the right
\begin{document}

\title{Nano-fabricated solid immersion lenses registered to single emitters in diamond}

\author{L. Marseglia}
\email[]{Luca.Marseglia@bristol.ac.uk}
%\homepage[]{Your web page}
%\thanks{}
%\altaffiliation{}
%\affiliation{University of Bristol}
%

\author{J. P. Hadden}
\author{A. C. Stanley-Clarke}
\author{J. P. Harrison}
\author{B. Patton}
\author{Y.-L. D. Ho}
\affiliation{Centre for Quantum Photonics, H. H. Wills Physics Laboratory \& Department of Electrical and Electronic Engineering, University of Bristol, Merchant Venturers Building, Woodland Road, Bristol, BS8 1UB, UK}
\author{B. Naydenov}
\author{F. Jelezko}
\affiliation{3 Physikalisches Institut and Research Center SCoPE, Universit\"{a}t Stuttgart, Pfaffenwaldring 57, 70569 Stuttgart, Germany}
\author{J. Meijer}
\affiliation{Ruhr-Universit\"{a}t Bochum, Universit\"{a}tsstra{\ss}e 150, 44801, Bochum, Germany}
\author{P. R. Dolan}
\author{J. M. Smith}
\affiliation{Department of Materials, University of Oxford,Parks Road, Oxford, OX1 3PH, UK}
\author{J. G. Rarity}
\author{J. L. O'Brien}
\affiliation{Centre for Quantum Photonics, H. H. Wills Physics Laboratory \& Department of Electrical and Electronic
Engineering, University of Bristol, Merchant Venturers Building, Woodland Road, Bristol, BS8 1UB, UK}
% Collaboration name, if desired (requires use of superscriptaddress option in \documentclass).
% \noaffiliation is required (may also be used with the \author command).
%\collaboration{}
%\noaffiliation

\begin{abstract}We describe a technique for fabricating micro- and nano-structures incorporating fluorescent defects in diamond with a positional accuracy in the hundreds of nanometers. Using confocal fluorescence microscopy and focused ion beam (FIB) etching we initially locate a suitable defect with respect to registration marks on the diamond surface and then etch a structure using these coordinates. We demonstrate the technique here by etching an 8 $\mu$m diameter hemisphere positioned such that a single negatively charged nitrogen-vacancy defect lies at its origin. This type of structure increases the photon collection efficiency by removing refraction and aberration losses at the diamond-air interface. We make a direct comparison of the fluorescence photon count rate before and after fabrication and observe an 8-fold increase due to the presence of the hemisphere.
\end{abstract}

\maketitle
\begin{figure}[t!]
\centering
\includegraphics[width=8.5cm, height=8cm]{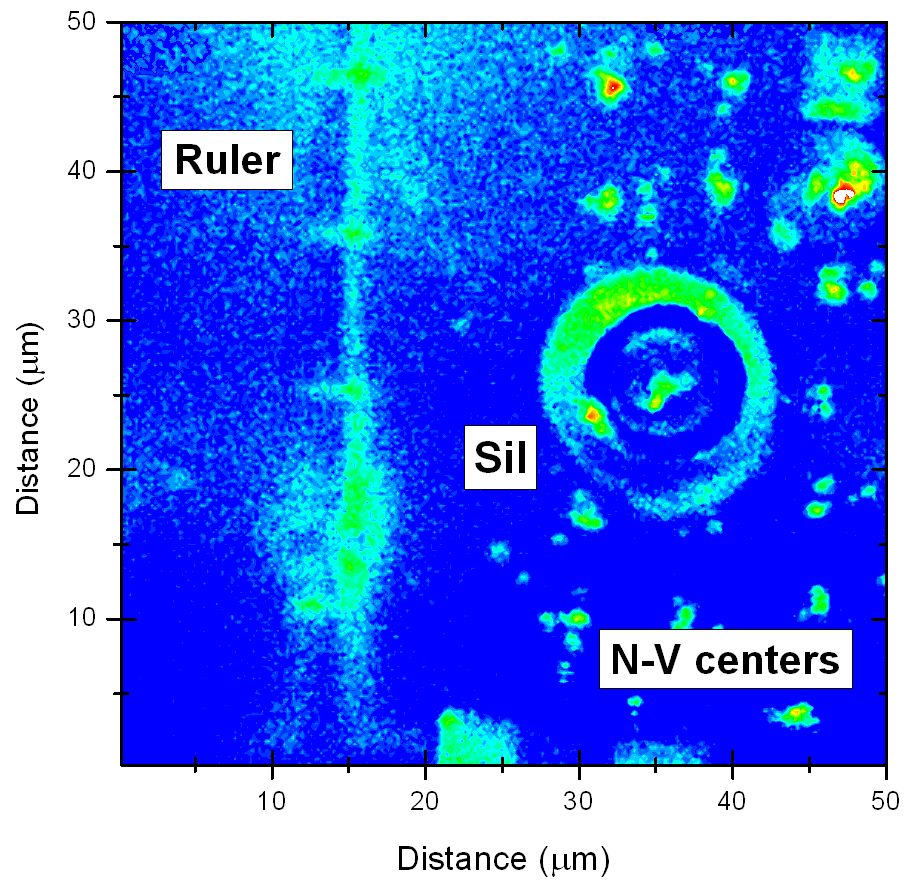}
 \caption{\scriptsize Confocal scanning image at a depth of 4 $\mu$m. Colour scale goes from blue ($5 \times 10^{3}$ cps) to Red ($2.5 \times 10^{4}$ cps). The array of implanted \nvsp centers and the SIL are both visible on the right of the image. Also visible, on the left, is the defocussed image of the ruler etched into the surface of the sample.}
\label{fig:MAP}
\end{figure}

\paragraph*{}Diamond has seen a resurgence of interest in its optically active defects (color centers), notably the nitrogen-vacancy (NV) center and most recently, various chromium-associated centers \cite{Aharonovich10}, all of which show promise as single photon sources\cite{Beveratos02b,Barnes02,AmpemLassen09}. Additionally, the negatively charged nitrogen-vacancy color center \nvsp is a spin active defect with a long spin lifetime at room temperature\cite{vanOortt98,Epstein05}. The center consists of a substitutional nitrogen atom adjacent to a carbon atom vacancy \cite{Davies76}. Experiments \cite{vanOortt98,Lenef96} have shown that the 637 nm direct optical recombination (zero phonon line, ZPL) is a multiplet of spin preserving transitions, allowing direct optical access to the spin state of the \nvsp center. The corresponding ratio of optical to spin lifetimes is in excess of 1 to $10^{4}$, thus fulfilling one of the Di Vincenzo criteria \cite{DiVincenzo96}. A second criterion is addressed by the presence of a spin dependent non radiative recombination pathway which leads to polarization of the spin in the ground state and therefore can be used as a spin initialization procedure\cite{Dutt08}. The ground state spin of the \nvsp center modulates the emission intensity, providing a mechanism for all-optical spin read-out schemes\cite{Balasubramanian09,Childress06,Hanson06}. Additionally, the non degeneracy of the ground state multiplet allows the use of spectrally high resolution resonant excitation as a direct probe of the spin state.
\paragraph*{}In order to use the spin state in a quantum information processing (QIP) device, or in quantum enhanced metrology (QEM), it is necessary to be able to use, the information of the spin state before the spin has decohered.
One of the current rate limiting steps for the development of diamond-based technology is related to the efficiency with which photons can be coupled from the optically active defect into the collection mode, or analysis channel, of the QIP or QEM device. The high refractive index mismatch at any diamond-air interfaces limits the solid angle from which photons can be collected due to refraction and adds additional Fresnel losses and strong spherical aberration which must be corrected for in the subsequent optical design. One solution is to use defects embedded in nanocrystalline diamond of a size much smaller than the optical wavelength. While this increases the coupling efficiency to the free-space modes, the \nvsps  experience high strain and the presence of nearby surface trap states which both cause unwanted mixing of the spin spates and provide an additional dephasing channel to the spin. It would therefore be preferable to use defects embedded in a meso- or macro-scopic diamond sample and to directly improve the coupling efficiency from a planar surface by fabricating structures such as solid immersion lenses (SILs) \cite{Hadden10,Jelezko10} or photonic wire micro-cavities \cite{Babinec10}.
\paragraph*{}Hemispherical SILs are interesting structures because they geometrically avoid any refraction at the diamond-air interface; if we consider a color center placed at the origin of a hemisperical diamond surface the fluorescent light coming out is perpendicular to the surface over the full $2\pi$ solid angle of the hemisphere. Recently\cite{Hadden10} we reported an increase by a factor of 10 in the saturated count-rate from a single \nvsp center using such structures. A focused ion beam (FIB) milling technique was used to etch hemispheres of 5 $\mu$m diameter into the surface of synthetic diamond material that had a moderate density of as-grown \nvsp centers. A number of these SILs were found to contain a single \nvsp center and, comparing the fluorescence intensity with the typical intensity of single \nvsps in an un-etched region, up to an order of magnitude improvement in photon collection was measured. However, the success achieved in this sample can be attributed to a combination of wise material choice (that is, a sufficiently high density of single centers giving a high probability of a defect being located within the central volume of a SIL) and the acceptance that a large fraction of the SILs fabricated may be empty of \nvsp centers. As such, we etched as many SILs as time permitted and then examined them to find one suitably located about an \nvsp center. It is clear that for scalability purposes, a different approach is required.
\paragraph*{}
In this paper we demonstrate a protocol for fabricating SILs incorporating single \nvsp centers in diamond in a pre-determined, rather than post-selected, way. Registration marks on the diamond surface that can be identified in our confocal fluorescence microscope allowed us to initially locate and then characterize particular \nvsp centers in a robust and repeatable way. These registration marks, also visible in the FIB system, then allowed us to etch a SIL with a single \nvsp center located precisely at the origin of the hemisphere. This approach eliminates the randomness of our previous approach and, importantly, allows us to make a direct comparison of the same \nvsp center before and after the FIB-etching process.
\paragraph*{}
The sample used in this work was a 5 x 5 x 1 mm single crystal type IIa diamond grown by chemical vapor deposition (CVD) containing substitutional nitrogen impurities below the detection limit (Element Six). Nitrogen implantation was performed over a total area of $\sim$1.6 x 0.2 mm in a grid pattern with 10 $\mu$m separation between implanted spots. An energy of 6 MeV was used so that the resulting implantation depth was $\sim$4 $\mu$m. The implanted area received an average nitrogen dose of 10 ions per implanted spot. Following implantation, the sample was annealed at 800$^{\circ}$C for 2 hours. At this temperature, carbon-vacancies (created by the ion implantation) become mobile in the diamond lattice and have some probability of forming \nvsp centers with the implanted nitrogen atoms.
\paragraph*{}In order to observe the array of implanted \nvsp centers (Fig.\ref{fig:MAP}) we used a confocal fluorescence microscope system in the reflection geometry, with excitation by a frequency-doubled Nd:YAG laser ($\lambda$=532nm) focused by an NA=0.9 microscope objective, giving a diffraction limited spatial resolution. The laser-induced fluorescence collected is focused onto the input of a single mode fibre which serves as the confocal aperture. Before collection, the fluorescence passes through two filters in order to block any reflected/scattered laser light and the first- and second-order Raman signal from the bulk diamond.

\begin{figure}[t!]
\centering
\includegraphics[width=8.5cm, height=8cm]{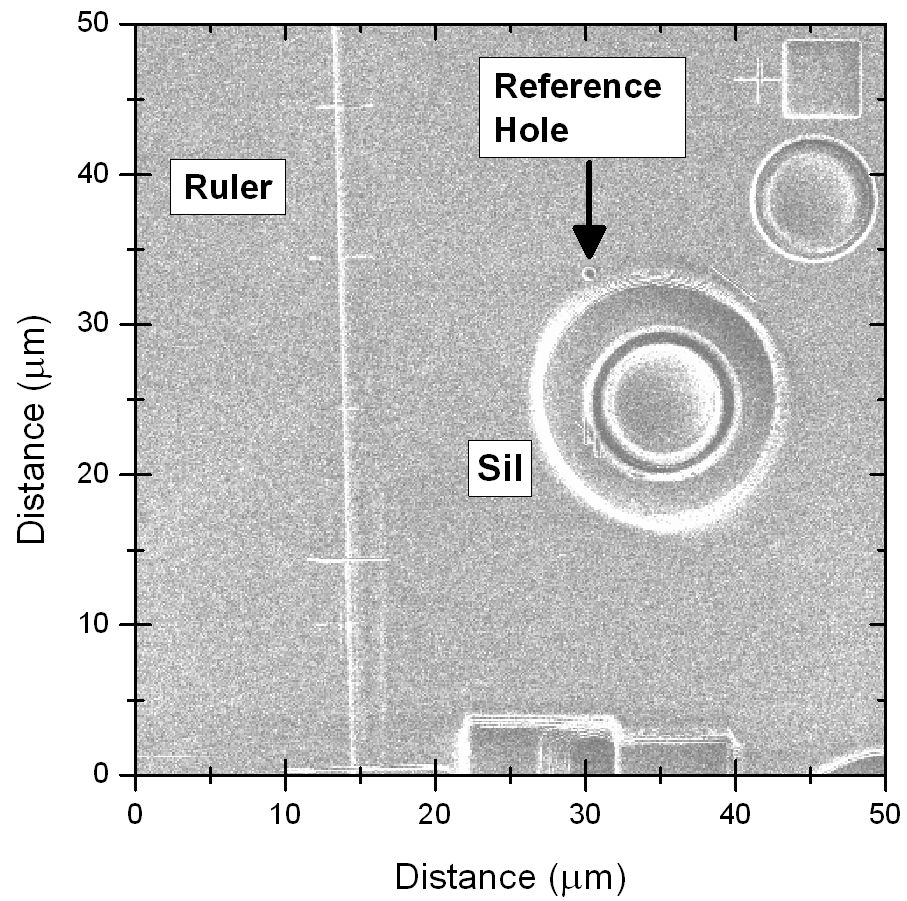}
 \caption{\scriptsize FIB image of the processed area on our sample. Highlighted is the etched ruler used for coarse positioning, the reference hole drilled for accurate location of the defect of interest, and the SIL described in the text. The structures in the top right of the image are test objects for calibration of the etching process.}
\label{fig:REFERENCE}
\end{figure}

The sample is mounted on a piezo-electric x-y-z stage with a scan range of ~80 $\mu$m for each axis, which is in turn mounted on a mechanical x-y-z stage for coarse positioning. In scanning mode, the fluorescence is detected by a single photon avalanche diode (SPAD) as the sample is scanned by the piezo stage and the resulting signal intensity is plotted to build up a fluorescence image. In analysis mode, the stage is static and the fluorescence signal can be directed either to an optical spectrometer for spectral analysis, or to two SPADs in a Hanbury-Brown and Twiss configuration to measure the second order intensity correlation function ${g}^{2}(\tau)$.
\paragraph*{}
Having a suitable sample, we next searched for single \nvsp centers by optically scanning across the entire implanted array. Once we have catalogued any single \nvsp centers we need to be able to precisely locate their position in order to correctly fabricate a SIL.
\paragraph*{}
We used a 30 keV gallium focused ion beam system (see our previous work\cite{Hadden10} for details) for fabrication. The FIB can also be used for imaging by rastering the ion beam at a low current and detecting the secondary electron emission. As diamond is dielectric and will build up charge during the FIB process, resulting in degradation of the FIB accuracy, a thin layer of platinum was deposited on the diamond surface before etching to minimize these effects.
\paragraph*{}
In order to etch structures precisely over color centers its necessary to be able to map between the fluorescence images of the NV centers and the FIB images of the diamond surface. To do this we etch registration marks that are identifiable in fluorescence using the confocal microscope. The position of the implanted region within the overall diamond slab was known with an accuracy of tens of microns, so the initial step in correlating the optical and FIB images of the sample was to use the FIB to etch a ``ruler" just outside the implanted region. We are then able to search for and catalogue the position of single NV- centers relative to the ruler using the confocal microscope(see Fig.\ref{fig:MAP}).
\paragraph*{}
A cautious approach was adopted because the ion beam must also be used to image the sample and it was unclear if damage and/or implantation from the gallium, even at low flux (but not low energy), might affect the implanted single \nvsp centers or contribute additional unwanted fluorescence.Great care was taken to clean the diamond before each step in the fabrication process, primarily to remove surface gallium, platinum, sp$^{2}$ carbon and organic contaminants. The cleaning procedure involved used acetone to remove dust-like contaminants, followed by an isopropanol clean to remove the acetone and a distilled water rinse for the isopropanol. An acid cleaning is then performed in 100 ml of sulphuric acid and 5 mg of potassium nitrate for 20 minutes at 200$^{\circ}$ C to remove more strongly bonded contaminants. After three iterations of the acid boil, the sample is rinsed thoroughly in deionised water. Having noted that imaging with gallium does not decrease the fluorescence of the \nvsp centers, we can then iterate the process to draw successively more precise reference marks with the FIB while continuing to minimize the area of the single \nvsp zone exposed to gallium irradiation.
\paragraph*{}
Once we are satisfied that we have a suitable crude correlation between the images from the FIB and the optical scans by using the etched ruler, we make an additional, and final, reference mark to precisely locate the single \nvsp by etching a hole of 300 nm radius $\sim$10 $\mu$m from it.
Fig.\ref{fig:REFERENCE} shows a FIB image of the FIB-etched ruler, the SIL and the final reference dot we etched close to the single \nvsp center.

\begin{figure}[t!]
\centering
\includegraphics[width=8.5cm, height=11cm]{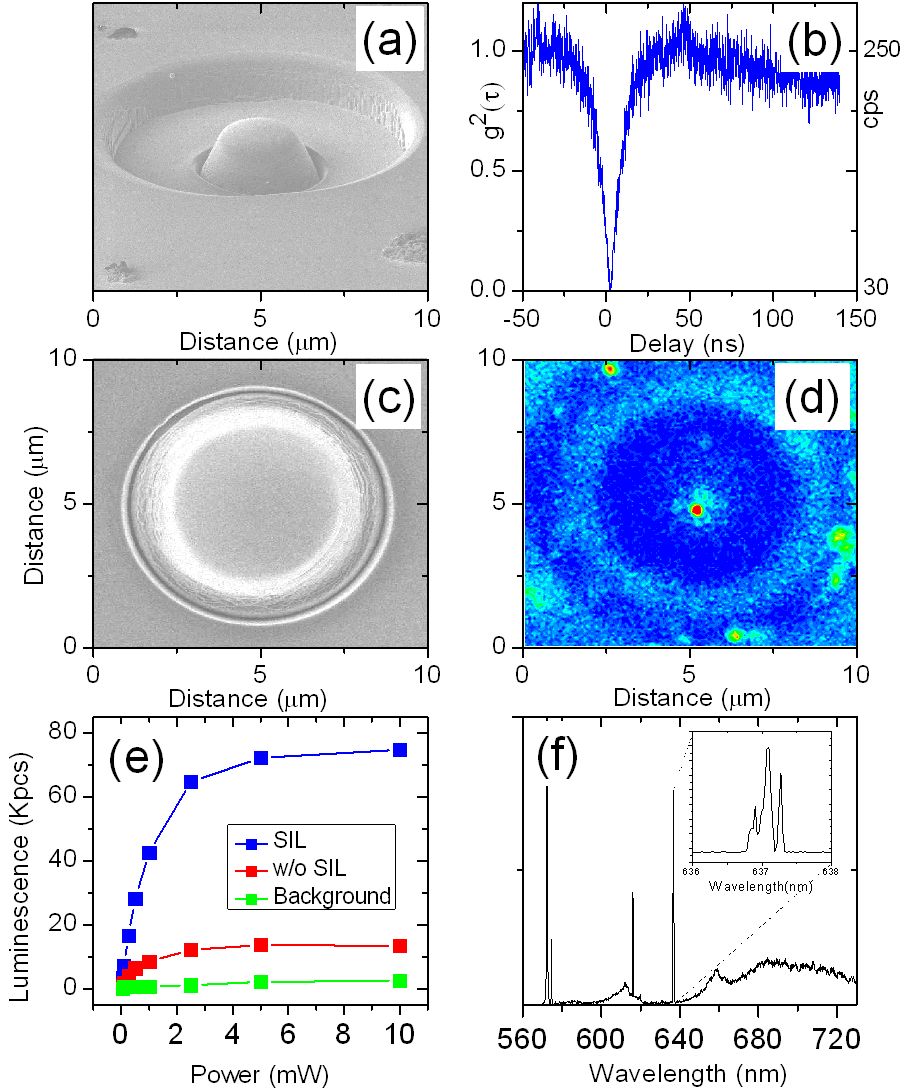}
 \caption{\scriptsize a) $45^{\circ}$ tilted FIB image of the etched SIL on the single \nvsp center,  b) second order autocorrelation function $g^{2}(\tau)$ (left axes:corrected coincidence counts,right axes:uncorrected coincidence counts)of the located single \nvsp center inside the fabricated SIL, taken at the excitation power of 0.1 mW c) FIB image of the etched SIL on the single \nvsp center, d) photoluminescence image of the SIL with the enhanced emission from the \nvsp center taken at temperature T=4.2K, e) Comparison of the different photoluminescence count rates as function of the laser intensity, f) Spectrum taken at temperature T=4.2K showing Raman line and the single \nvsp center.}
\label{fig:SIL}
\end{figure}

\paragraph*{}
The SIL structure itself was etched in two stages: at the beginning an annular trench of 20 $\mu$m outer diameter, inner diameter of 8 $\mu$m and 4 $\mu$m deep, and subsequently an 8 $\mu$m diameter hemisphere centered in the trench (i.e. the top of the SIL is now level with the surface of the bulk diamond). In both stages, a test structure was initially etched outside of the implanted zone. This is because, even with the platinum deposition, there are still charging effects that cause the FIB to be deflected. To compensate for this problem, we attempt to etch our test structure (trench or hemisphere) with its center at known coordinates with reference to the FIB field of view. Then we measure the difference between these coordinates and the actual center of the etched structure. We can use this offset value when we etch the real structure to place the color center at the appropriate position within the FIB field of view.
\paragraph*{}
The results of this SIL fabrication process are shown in Fig.\ref{fig:SIL}a, and Fig.\ref{fig:SIL}c. In Fig.\ref{fig:SIL}d we show a confocal image of the NV with additional structure in the background noise showing the geometry of the SIL. This image shows our ability to center the NV on the SIL, it is also worth noting that the SIL imparts an effective magnification equivalent to the refractive index n =2.4 of the diamond, thus making the apparent radial centering error of 500 nm translate to an actual positioning error of 210 nm. This is within the calculated error \cite{Castelletto11} for which the SIL operates at greater than 90\% of the maximum efficiency. At the center of the trench a brightly fluorescent spot can be seen clearly. The second order autocorrelation function measurement, shown in Fig.\ref{fig:SIL}b, again confirms single \nvsp emission. We normalized and corrected the data for background as described by Beveratos et al. \cite{Beveratos02b}. The photon count rate as a function of excitation power was measured before and after etching the SIL, allowing us to make a direct comparison of the saturation curves for the single \nvsp (Fig.\ref{fig:SIL}e). From this we see that the saturated photon count rate has increased by a factor of 8.
\paragraph*{}
We also performed spectroscopy of the defect at liquid helium temperatures to observe the effect of the SIL on the \nvsp. Fig.\ref{fig:SIL}f shows a typical spectrum. The sharp line at 574 nm is the first order Raman signal from the 532 nm excitation source and is accompanied by a broad second-order Raman peak centered at 615 nm. The NV$^{0}$ peak at 576nm is clearly visible, as is the \nvsp emission with its ZPL at 637 nm and the spectrally broad phonon-assisted emission to longer wavelengths. The inset to this figure shows a high resolution spectrum of the ZPL. The splitting evident is consistent with spectra from other implanted \nvsp centers and is likely due to state mixing in the excited state due to strain from vacancies generated in the implantation procedure and does not result from the FIB process.

\paragraph*{}Summarizing, we have demonstrated a technique that allowed us to precisely locate, with positional accuracy limited by the optical resolution and pointing accuracy, a single \nvsp center in diamond and fabricate a hemispherical SIL on it. The process is independent of the color center studied and will allow us in the future the possibility of creating SILs placed on different color centers, such as chromium centers. The same technique could be also used to create different structures, such as photonic crystal cavities, etched exactly onto the place of the emitter allowing us to measure the coupling between the \nvsp and the photonic crystal cavity. This type of structure would be suitable, for example, to perform quantum non demolition spin readout measurement as we have recently shown theoretically in another work \cite{Young09}, where we demonstrated that spin readout with a small number of photons could be achieved by placing the \nvsp center in a subwavelength scale micro-cavity with a moderate Q-factor. The technique we reported here opens further routes to the fabrication of nano-scale and micro-scale structures, precisely aligned to single defects for a wide range of applications.

\bibliography{bibliography}

\end{document}